\documentclass[sts]{imsart}

\RequirePackage{amsthm,amsmath,amsfonts,amssymb}
\RequirePackage[authoryear]{natbib}
\RequirePackage[colorlinks,citecolor=blue,urlcolor=blue]{hyperref}
\RequirePackage{graphicx}

\startlocaldefs
\usepackage{amsfonts}
\usepackage{lscape}
\usepackage{color}

\startlocaldefs
\definecolor{DarkO}{cmyk}{0.0, 0.43, 0.90, 0.13}

\endlocaldefs

\begin{document}

\begin{frontmatter}
\title{Computing Bayes: From Then `Til Now\thanksref{T1}}
\thankstext{T1}{Material
that appeared in an earlier paper, entitled: `Computing Bayes: Bayesian
Computation from 1763 to the 21st Century' (\citealp{martin2020computing}),
now appears (in amended form) in two separate papers: the current one, in
which a historical overview of and timeline for all computational
developments is presented; and a second one (\citealp{martin2021approximating}), 
which provides a detailed review of 21st
century approximate methods. Martin and Frazier have been supported by
Australian Research Council Discovery Grants DP170100729 and DP200101414,
and the Australian Centre of Excellence in Mathematics and Statistics.
Frazier has also been supported by Australian Research Council Discovery
Early Career Researcher Award DE200101070. Robert has been partly supported
by a senior chair (2016-2021) from l'Institut Universitaire de France and by
a Prairie chair from the Agence Nationale de la Recherche
(ANR-19-P3IA-0001).}

\begin{aug}
\author[A]{\fnms{Gael M.}~\snm{Martin}\thanksref{t1}\ead[label=e1]{gael.martin@monash.edu.}},
\author[B]{\fnms{David T.}~\snm{Frazier}\ead[label=e2]{david.frazier@monash.edu}} 
\and
\author[C]{\fnms{Christian P.}~\snm{Robert}\ead[label=e3]{christian.robert@dauphine.psl.edu}}
\thankstext{t1}{Corresponding author}
\address[A]{Monash University, Melbourne\printead[presep={\ }]{e1}.}
\address[B]{Monash University, Melbourne\printead[presep={\ }]{e2}.}
\address[C]{Université Paris Dauphine and University of Warwick\printead[presep={\ }]{e3}.}

\end{aug}

\begin{abstract}
This paper takes the reader on a journey through the
history of Bayesian computation, from the 18th century to the present day.
Beginning with the one-dimensional integral first confronted by Bayes in
1763, we highlight the key contributions of: Laplace, Metropolis (and,
importantly, his co-authors!), Hammersley and Handscomb, and Hastings, all
of which set the foundations for the computational revolution in the late
20th century --- led, primarily, by Markov chain Monte Carlo (MCMC)
algorithms. A very short outline of 21st century computational methods ---
including pseudo-marginal MCMC, Hamiltonian Monte Carlo, sequential Monte
Carlo, and the various `approximate' methods --- completes the paper.
\end{abstract}

\begin{keyword}
\kwd{History of Bayesian computation}\kwd{ Laplace approximation}\kwd{
Metropolis-Hastings algorithm}\kwd{ importance sampling}\kwd{ Markov chain Monte
Carlo}\kwd{ pseudo-marginal methods}\kwd{ Hamiltonian Monte Carlo}\kwd{ sequential Monte
Carlo}\kwd{ approximate Bayesian methods}
\end{keyword}

\end{frontmatter}


\section{The Beginning}

\textit{December 23 1763, London}: Richard Price reads to the Royal Society
a paper penned by a past Fellow, the late Thomas Bayes:\smallskip

\begin{quote}
`\textit{An Essay Towards Solving a Problem in the Doctrine of Chances.}%
'\smallskip
\end{quote}

\noindent With that {reading}, the concept of `inverse probability' --- 
\textit{Bayesian inference} as we know it now --- has its first public
airing.

To our modern eyes, the problem tackled by Bayes in his essay is a simple
one: If one performs $n$ independent Bernoulli trials, with a probability, $%
\theta $, of success on each trial, what is the probability \textbf{---}
given $n$ outcomes \textbf{---} of $\theta $ lying between two values, $a$
and $b$? The answer Bayes offered is equally simple to re-cast in modern
terminology. Define $Y_{i}|\theta \sim i.i.d.$ $Bernoulli(\theta )$, $%
i=1,2,...,n$; record the observed sequence of successes ($Y_{i}=1$) and
failures ($Y_{i}=0$) as $\mathbf{y}=(y_{1},y_{2},...,y_{n})^{\prime }$;
denote by $p(\mathbf{y}|\theta )$ the likelihood function for $\theta $; and
invoke a Uniform prior, $p(\theta )$, on the interval $(0,1)$. Bayes sought:%
\begin{equation}
\mathbb{P}(a<\theta <b|\mathbf{y})=\int\limits_{a}^{b}p(\theta \mathbf{|y}%
)d\theta ,  \label{Bayes_prob}
\end{equation}%
where $p(\theta \mathbf{|y})$ denotes the posterior probability density
function (pdf) for $\theta $,%
\begin{equation}
p(\theta \mathbf{|y})=\frac{p(\mathbf{y}|\theta )p(\theta )}{p(\mathbf{y})},
\label{posterior_pdf}
\end{equation}%
$p(\mathbf{y})=\int_{0}^{1}p(\mathbf{y}|\theta )p(\theta )d\theta $ defines
the marginal likelihood, and the scale factor $\left[ p(\mathbf{y})\right]
^{-1}$ in (\ref{posterior_pdf}) ensures that $p(\theta |\mathbf{y})$
integrates to one. Given the Bernoulli assumption for $Y|\theta $, the
Uniform prior on $\theta $, and defining $x=\Sigma _{i=1}^{n}y_{i}$, $%
p(\theta \mathbf{|y})$ has a closed-form representation as the Beta pdf, 
\begin{equation}
p(\theta \mathbf{|y})=\left[ B(x+1,n-x+1)\right] ^{-1}\theta ^{x}(1-\theta
)^{n-x},  \label{beta_pdf}
\end{equation}%
where $B(x+1,n-x+1)=\Gamma (x+1)\Gamma (n-x+1)/\Gamma
(n+2)=\int_{0}^{1}\theta ^{x}(1-\theta )^{n-x}d\theta $ is the Beta
function, and $\Gamma (x)$ is the Gamma function. Bayesian inference \textbf{%
---} namely, quantification of uncertainty about an unknown $\theta $,
conditioned on known data, $\mathbf{y}$ \textbf{---} thus first emerges as
the analytical solution to a particular inverse probability problem.%
\footnote{%
Bayes cast his problem in physical terms: as one in which balls were rolled
across a square table, or plane. Over time his pictorial representation of
the problem has come to be viewed as a `billiard table', despite Bayes
making no overt reference to such an item in his essay. For this, and other
historical anecdotes, see \cite{stigler:1986} and \cite{fienberg:2006}.}

Bayes, however, did not seek the pdf in (\ref{beta_pdf}) \textit{per se}.
Rather, he wished to evaluate the probability in (\ref{Bayes_prob}) which,
for either $a\neq 0$ or $b\neq 1$, {involved evaluation of the }incomplete{\
Beta function. Except for when either }$x$ {or }$(n-x)$ was {small, a
closed-form solution to }(\ref{Bayes_prob}) {eluded Bayes. }Hence, despite
the analytical availability of $p(\theta \mathbf{|y})$ via (\ref%
{posterior_pdf}) \textbf{---} `Bayes' rule' {as it is now known} \textbf{---}
{the} quantity that was {of interest to Bayes }needed to be estimated, {or }%
\textit{computed. }The quest for a computational {solution to a} Bayesian
problem was\ thus born.

\section{Preliminaries}\label{general}

\subsection{Why Do We Need Numerical Computation?}

Bayes' probability of interest in (\ref{Bayes_prob}) can, of course, be
expressed as a posterior expectation, $\mathbb{E}(\mathbb{I}_{[a,b]}\mathbf{|y)}=\int_{\Theta }\mathbb{I}_{[a,b]}p(\theta \mathbf{|y})d\theta ,$ where $%
\mathbb{I}_{[a,b]}$ is the indicator function on the interval $[a,b]$.
Generalizing at this point to any problem with unknown $\boldsymbol{\theta }%
=(\theta _{1},\theta _{2},...,\theta _{p})^{\prime }\in $ ${\Theta }$ and
joint posterior pdf $p(\boldsymbol{\theta }\mathbf{|y})$, most Bayesian
quantities of interest are posterior\textit{\ }expectations of some function 
$g(\boldsymbol{\theta })$ and, hence, can be expressed as,

\begin{equation}
\mathbb{E}(g(\boldsymbol{\theta }\mathbf{)|y)}=\int_{{\Theta }}g(\boldsymbol{%
\theta }\mathbf{)}p(\boldsymbol{\theta }\mathbf{|y})d\boldsymbol{\theta }%
\mathbf{.}  \label{gen_expect}
\end{equation}%
In addition to posterior probabilities like that of Bayes, familiar examples
include posterior moments, marginal posterior densities and moments,
predictive densities and posterior expectations of loss functions. Moreover,
conditioning explicitly on the assumed model $\mathcal{M},$ the marginal
likelihood of the model is the \textit{prior }expectation,%
\begin{equation}
p(\mathbf{y}|\mathcal{M})=\int_{{\Theta }}p(\mathbf{y}|\boldsymbol{\theta },%
\mathcal{M})p(\boldsymbol{\theta }|\mathcal{M})d\boldsymbol{\theta }.
\label{gen_expect_prior}
\end{equation}%
The ratio of (\ref{gen_expect_prior}) to the corresponding quantity for an
alternative model defines the \textit{Bayes factor }for use in model choice.
(See \citealp{berger:1985}, \citealp{koop2003bayesian}, \citealp{Geweke2005}%
, and \citealp{robert:2007}, for textbook expositions).

The key point to note is that analytical solutions to (\ref{gen_expect}) and
(\ref{gen_expect_prior}) are usually unavailable. Indeed, Bayes' original
problem highlights that a solution to (\ref{gen_expect}) can elude us 
\textit{even} when the posterior pdf itself has a closed form. Typically,
the posterior is known only up to a constant of proportionality, as 
\begin{equation}
p(\boldsymbol{\theta }\mathbf{|y})\propto p(\mathbf{y}|\boldsymbol{\theta }%
)p(\boldsymbol{\theta }),  \label{Bayes_proport}
\end{equation}%
exceptions to this including when $p(\mathbf{y}|\boldsymbol{\theta })$ is
from the exponential family, and either a natural conjugate, or convenient
noninformative prior is adopted {(}as in Bayes'{\ problem)}. Knowledge of $p(%
\boldsymbol{\theta }\mathbf{|y})$ only up to the integrating constant\textit{%
\ }implies a lack of closed-form solution for (\ref{gen_expect}), no matter
what the form of $g(\boldsymbol{\theta })$. A lack of knowledge of the
integrating constant automatically implies that the marginal likelihood for
the assumed model in (\ref{gen_expect_prior}) is unavailable. Situations
where the likelihood function itself does not have a closed form render the
analytical solution of (\ref{gen_expect}) and (\ref{gen_expect_prior}) an
even more distant dream. In short, the implementation of all forms of
Bayesian analysis relies heavily on numerical computation.

\subsection{The Structure of this Review}

In writing this review we have made two key decisions: \textit{i)} to
describe all methods using a common notation; and \textit{ii)} to place the
evolution of computational methods in a historical context, making use of
pictorial timelines in the process. In so doing, we are able to present a
coherent chronological narrative about Bayesian computation up to the
present day. Specifically, all methods can be seen to be, in essence,
attempting to compute integrals like (\ref{gen_expect}) and (\ref%
{gen_expect_prior}); the use of a common notation makes that clear. However,
important details of those integrals have changed over time: the dimension
of $\boldsymbol{\theta }$ (i.e. the number of$\ $`unknowns'), the dimension
of $\mathbf{y}$ (i.e. the `size' of the data), the nature of the integrand
itself and, most critically, the available computing technology. Computation
has evolved accordingly, and the chronological ordering helps make sense of
that evolution. Hence, whilst computational methods can \textbf{---} and
often are \textbf{---} grouped according the category into which they fall,
e.g. \textit{deterministic, simulation-based, approximate }etc., the
over-arching structure that we adopt here is one of chronology, as
understanding \textit{when} a computational method has appeared aids in the
appreciation of \textit{why}, and \textit{how}.

We begin, in Section \ref{timeline}, by returning to Bayes' integral in (\ref%
{Bayes_prob}), briefly reiterating the nature of the particular
computational problem it presented. We then use this as a springboard for
pinpointing four particular points in time during the two centuries (or so)
subsequent to 1763: 1774, 1953, 1964 and 1970. These time points correspond,
in turn, to four publications \textbf{---} by Laplace, Metropolis \textit{et
al., }Hammersley and Handscomb, and Hastings, respectively \textbf{---} in
which computational methods that produce estimates of integrals like that of
Bayes, were proposed. Whilst only the method of Laplace was explicitly set
within the context of inverse probability (or Bayesian inference), all four
methods of computing integrals can be viewed as harbingers of what was to
come in Bayesian computation \textit{per se}.

In Section \ref{70s}, we look at Bayesian computation in the late 20th
century, during which time the inexorable rise in the speed, and
accessibility, of computers led to the pre-eminence of \textit{%
simulation-based }computation. Whilst important new developments arose that
exploited the principles of importance sampling (\citealp{kloek1978bayesian}%
; \citealp{geweke:1989}; \citealp{gordon:salmon:smith:1993}), the
computational engine was well and truly fuelled by Markov chain Monte Carlo
(MCMC) algorithms, allied with the concept of `data augmentation'; with \cite%
{besag:1974}, \cite{geman:1984}, \cite{tanner87} and \cite{gelfand:smith90}
being seminal contributions.

Section \ref{second} then provides a potted summary of a `second
computational revolution' in the 21st century --- born in response to the
increased complexity, and size, of the empirical problems being analyzed via
Bayesian means. We begin by briefly noting the important modifications and
refinements of MCMC that have occurred since its initial appearance,
including pseudo-marginal methods, Hamiltonian up-dates, adaptive sampling,
and coupling; plus highlighting the renaissance of importance sampling that
has occurred under the auspices of sequential Monte Carlo (SMC) methods. We
then very briefly outline the main \textit{approximate }methods that evolved
to tackle intractable problems: approximate Bayesian computation (ABC),
Bayesian synthetic likelihood (BSL), variational Bayes (VB) and integrated
nested Laplace approximation\textit{\ }(INLA).

Throughout the paper we explicitly reference only computational solutions%
\textbf{\ }to the posterior expectation in (\ref{gen_expect}) which, of
course, covers the use of computation in Bayesian prediction, via the
appropriate choice of $g(\boldsymbol{\theta }\mathbf{)}$. For coverage of
the use of computational methods to solve the prior expectation in (\ref%
{gen_expect_prior}) we refer the reader to \cite{ARDIA2012} and \cite%
{llorente2021marginal}.

Finally, in order to provide the reader with a quick `snapshot' of the key
personages involved in the evolution of computation (and the relevant
seminal publications), Figures 1, 2 and 3 provide pictorial timelines
coinciding approximately with the time periods covered in Sections \ref%
{timeline}, \ref{70s} and \ref{second}, respectively.

\section{Some Early Chronological Signposts: 1763-1970}\label{timeline}

\begin{figure}[th]
\centering
\includegraphics[scale=0.6]{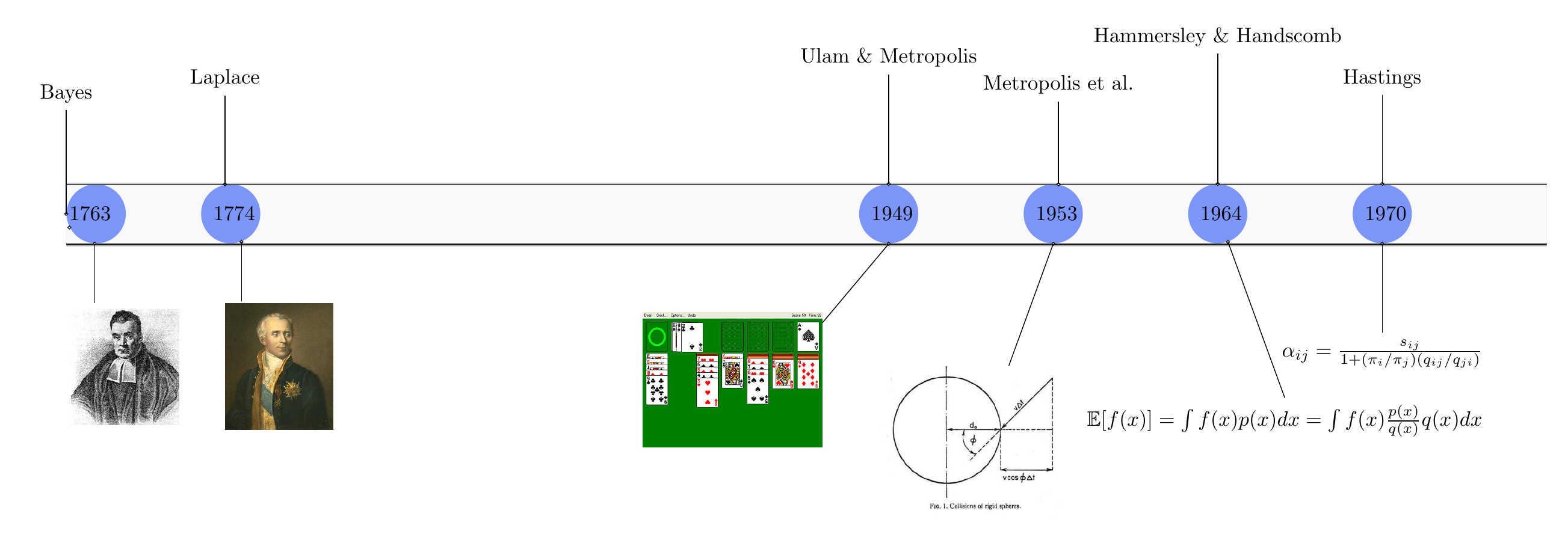}
\caption{From left to right: \textit{1)} The only known portrait (albeit
unauthenticated) of the Reverend Thomas Bayes (1702-1761); \textit{2)} A
portrait of Pierre-Simon Laplace (1949-1827); \textit{3)} A snapshot of the
game of Solitaire. Stanislaw Ulam speculated that the proportion of wins in
repeated random games played by a computer could be used as an estimate of
the probability of winning the game --- thereby giving birth to the idea of
estimation by Monte Carlo simulation; \textit{4)} A diagram illustrating the
`collisions of rigid spheres' used in the description of the behaviour of
interacting particles in \protect\cite{metropolis:1953}; \textit{5)}
Equivalent representations of the expectation of a function of a random
variable $x$, $f(x)$. This equivalence underlies the use of repeated draws
of $x$ from $q(x)$ to produce an `importance sampling' estimate of $\mathbb{E%
}[f(x)]$; \textit{6)} The quantity used in \protect\cite{hastings:1970} to
define a Markov chain, based on the transition matrix $\boldsymbol{Q}%
=\{q_{ij}\}$, with invariant distribution $\boldsymbol{\protect\pi }=(%
\protect\pi _{0},\protect\pi _{1},...,\protect\pi _{S})$ over states $%
0,1,...,S.$}
\end{figure}

\subsection{1763: Bayes' Integral}

Bayes' desire was to {evaluate} the probability in (\ref{Bayes_prob}). As
noted {above}, {for either} $a\neq 0$ {or} $b\neq 1$, {this required
evaluation of the incomplete Beta function. For either }$x$ {or }$(n-x)$ {%
small}, Bayes {proposed} a Binomial expansion and term-by-term integration {%
to give an exact solution (his `Rule 1')}. {However, for }$x$ {and }$(n-x)$ {%
both large, this approach was infeasible: prompting Bayes (and,
subsequently, Price himself; \citealp{price:1764})} to {resort to producing
upper and lower bounds for (\ref{Bayes_prob}) using quadrature}. Indeed, {%
Stigler (1986a) speculates} that the inability to produce {an approximation
to (\ref{Bayes_prob})} that was sufficiently accurate may explain Bayes'
reluctance to publish his work and, {perhaps}, the lack of attention it
received subsequent to its (posthumous) presentation {by Price }in 1763%
\textbf{\ }and publication the following year {in \cite{bayes:1764}}.%
\footnote{%
On November 10, 1763, Price sent an edited and annotated version of Bayes'
essay to the {Secretary of the} Royal Society, with his own Appendix added.
Price read the essay to the Society on December 23, as noted earlier. The
essay and appendix were subsequently published in 1764, in the \textit{%
Philosophical Transactions of the Royal Society of London}. The front matter
of the issue appears here:
https://royalsocietypublishing.org/cms/asset/f005dd95-c0f8-45b2-8347-0296a93c4272/front.pdf. The publication has been reprinted since, including in 
\cite{Bayes:Biometrika1958}, with a biographical note by G.A. Barnard.
Further historical detail on the important role played by Price in the
dissemination of Bayes' ideas can be found in \cite{Hooper:2013} and \cite%
{stigler2018}. As the submission of Bayes' essay by Price, and his
presentation to the Royal Society occurred in 1763, and Volume 53 of the 
\textit{Philosophical Transactions} in which the essay appears\ is `For the
Year 1763', Bayes' essay is often dated 1763. We follow \cite{stigler:1986}
in using the actual publication date of 1764.}

Whilst the integral that Bayes wished to compute was a very particular one,
it was representative of the general hurdle that needed to be overcome if
the principle of inverse probability were to be a useful practical tool. In
brief, inference\textit{\ }about $\theta $ was expressed in probabilistic
terms and, hence, required either the direct computation of probability
intervals, or the computation of distributional moments of some sort.
Ironically, the choice of the Bernoulli model, possibly the simplest process
for generating data `forward' (conditional on $\theta $) that Bayes could
have assumed, exacerbated this problem, given that the `inversion' problem
does not possess the simplicity of the generative problem. What was required
was a solution that was, in large measure, workable no matter what the
nature of the generative model, and the first solution came via the 1774 
\textit{`M\'{e}moire sur la probabilit\'{e} des causes par les \'{e}v\'{e}%
nemens'} by Pierre Laplace.

\subsection{1774: Laplace and His Method of Asymptotic Approximation}\label{Laplace}

Laplace envisaged an experiment in which $n$ tickets were drawn with
replacement from an urn containing a given proportion of white and black
tickets. Recasting his analysis in our notation, $\theta $ is the
probability of drawing a white ticket, $\mathbf{y}=(y_{1},y_{2},...,y_{n})^{%
\prime }$ denotes the sequence of observed white tickets ($Y=1$) and black
tickets ($Y=0$) associated with $n$ independent draws of the random variable 
$Y|\theta \sim \text{Bernoulli}(\theta )$, and $x=$ $\Sigma _{i=1}^{n}y_{i}$
is the number of white tickets drawn. Laplace's aim was to show that, for
arbitrary $w$: $\mathbb{P}(\left\vert \frac{x}{n}-\theta \right\vert <w|%
\mathbf{y})=\mathbb{P}(\frac{x}{n}-w<\theta <\frac{x}{n}+w|\mathbf{y}%
)\rightarrow 1$ as $n\rightarrow \infty .$ That is, Laplace wished to
demonstrate \textit{posterior consistency}: concentration of the posterior
onto the true proportion of white tickets in the urn, $\theta _{0}=\underset{%
n\rightarrow \infty }{\lim }\frac{x}{n}$. Along the way, however, he
stumbled upon the same problem as had Bayes: computing the following
probability of a Beta random variable, 
\begin{equation}
\mathbb{P}(a<\theta <b|\mathbf{y})=\dfrac{
\int_{a}^{b}\theta ^{x}(1-\theta )^{n-x}d\theta}
{B(x+1,n-x+1)} ,  \label{lp}
\end{equation}%
with $a=\frac{x}{n}-w\neq 0$ and $b=\frac{x}{n}+w\neq 1.$ Laplace's genius
(allied with the power of asymptotics!) was to recognize that the
exponential of the integrand in (\ref{lp}) has the bulk of its mass in the
region of its mode, as $n$ gets large, and that the integral can be computed
in closed form in this case. This enabled him to prove (in modern notation)
that $\mathbb{P}(\left\vert \theta -\theta _{0}\right\vert >w|\mathbf{y}%
)=o_{p}(1),$ where $p$ denotes the probability law of $Y|\theta _{0}.$

The route he took to this proof, however, involved approximating the Beta
posterior with a Normal distribution, which (under regularity) is an
approach that can be used to provide a large sample approximation of
virtually \textit{any} posterior probability.\ Specifically, expressing an
arbitrary posterior probability as%
\begin{equation}
\mathbb{P}(a<\theta <b|\mathbf{y})=\int\limits_{a}^{b}p(\theta \mathbf{|y}%
)d\theta =\int\limits_{a}^{b}\,e^{ nf(\theta )} d\theta ,
\label{la_prob}
\end{equation}%
where $f(\theta )=\log \left[ p(\theta \mathbf{|y})\right] /n$, a
second-order Taylor series approximation of $f(\theta )$ around its mode, $%
\widehat{\theta }$, yields the \textit{Laplace asymptotic approximation} 
\begin{equation}
\begin{array}{cl}
\mathbb{P}(a<\theta <b|\mathbf{y}) & \approx e^{ nf(\widehat{\theta 
})}\, \sqrt{2\pi}\sigma \{\Phi \lbrack \frac{b-\widehat{%
\theta }}{\sigma }]-\Phi \lbrack \frac{a-\widehat{\theta }}{\sigma }]\},%
\end{array}
\label{lap}
\end{equation}%
where $\sigma ^{2}=-[nf^{^{\prime \prime }}(\widehat{\theta })]^{-1}$ and $%
\Phi (.)$ denotes the standard Normal cumulative distribution function.%
\footnote{%
See \cite{tierney:kadane:1986} and Robert and Casella (2004) for further
elaboration; and \cite{ghosal1995convergence} and \cite{vandervaart:1998}
for more formal demonstrations of the conditions under which a posterior
distribution converges in probability to a Normal distribution, and the
so-called Bernstein-von Mises theorem --- the modern day version of
Laplace's approximation --- holds.}

With (\ref{lap}), Laplace had thus devised a general way of implementing
inverse probability: probabilistic statements about an unknown parameter, $%
\theta $, conditional on data generated from any (regular) model, could now
be made, at least up to an error of approximation. Whilst his focus was
solely on the computation of {a specific posterior} {probability}, and in a
single parameter setting, {his method} was eventually used to approximate
general posterior expectations of the form in (\ref{gen_expect}) (%
\citealp{lindley1980bayesian}; \citealp{tierney:kadane:1986}; %
\citealp{tierney:kass:kadane:1989}) and, indeed, applied as an integral
approximation method in its own right \citep{bruijn1961asymptotic}. The
approach also underpins the modern INLA technique to be mentioned in Section %
\ref{approx} \citep{rue:martino:chopin:2009}.\footnote{%
Stigler (1975, Section 2) states that he has found no documentary evidence
that Laplace's ideas on inverse probability, as presented in the 1774
publication, including his own statement of `Bayes' theorem' in (\ref%
{posterior_pdf}), were informed by Bayes' earlier ideas. See Stigler (1986a,
Chapter 3) for discussion of Laplace's later extensions of Bayes' theorem to
the case of a non-Uniform prior, and see \cite{stigler:1975}, \cite%
{stigler:1986}, {\cite{stigler:Laplace1774}} and \cite{fienberg:2006} {on
matters of attribution}. The first recorded reference to Bayes' prior claim
to inverse probability is in the preface, written by Condorcet, to Laplace's
later 1781 publication: `\textit{M\'{e}moire sur les probabilit\'{e}s}'.}

Meanwhile, it would take 170-odd years for the \textit{next} major advance
in the computation of probability integrals to occur; an advance that would
eventually transform the way in which problems in inverse probability could
be tackled. This development was based on a new form of thinking and,
critically, required a platform on which such thinking could operate:
namely, machines that could \textit{simulate }repeated\textit{\ }random
draws of $\boldsymbol{\theta }$ from $p(\boldsymbol{\theta }\mathbf{|y})$,
or from some representation thereof. Given a sufficient number of such
draws, and the correct use of them, an estimate of (\ref{gen_expect}) could
be produced that \textbf{---} unlike the Laplace approximation \textbf{---}
would be accurate for any sample size, $n$, and would require less
analytical input. This potential to accurately estimate (\ref{gen_expect})
for essentially any problem, and any given sample size, was the catalyst for
a flourishing of Bayesian inference in the late 20th century and beyond. The
1953 publication in the \textit{Journal of Chemical Physics }by Metropolis,
Rosenbluth, Rosenbluth, Teller and Teller: \textit{`Equation of State
Calculations by Fast Computing Machines'}, was a first major step in this
journey.\footnote{%
With reference to the mechanical simulation of a random variable, we
acknowledge the earlier 1870s' invention of the \textit{quincunx} by Francis
Galton. This machine used the random dispersion of metal shot to illustrate
(amongst other things) draws from a hierarchical Normal model and regression
to the mean. Its use can thus be viewed as the first illustration of the
conjugation of a Normal likelihood and a Normal prior. See \cite%
{stigler:1986} for more details, including Galton's graphical illustration
of his machine in a letter to his cousin (and Charles Darwin's son), George
Darwin.}

\subsection{1953: Monte Carlo Simulation and the Metropolis Algorithm}
\label{MC}

The convergence of the idea of simulating random draws from a probability
distribution, and the production of such draws by computing machines,
occurred in the scientific hothouse of the Los Alamos Laboratory, New
Mexico, in the 1940s and 1950s; the primary impetus being the need to
simulate physical processes, including neutrons in the fissile material in
atomic bombs. We refer the reader to \cite{liu01}, \cite{hitchcock:2003}, {%
\cite{Gub2005}} {and }\cite{robert:casella:2011} for reviews of this period,
including details of the various personalities who played a role therein.%
\footnote{%
We make particular mention here of John and Klara von Neumann, and Stanislav
Ulam, with the latter co-authoring the 1949 publication in the \textit{%
Journal of the American Statistical Association: `The Monte Carlo Method' }%
with\ Nicholas Metropolis. We also note the controversy concerning the
respective contributions of the five authors of the 1953 paper (who included
two married couples). {On this particular point, we refer the reader to the
informative 2005 article by Gubernatis, in which Marshall Rosenbluth gives a
bird's eye account of who did what, and when. The article brings to light
the important roles played by both Adriana Rosenbluth and Mici Teller.
Finally, we alert the reader to the proceedings of a symposium on the Monte
Carlo method held at the University of California in July, 1949, edited by
A.S. Householder of the Oak Ridge National Laboratory, Tennessee.
Researchers at the Oak Ridge laboratory were also involved in the early
exploration of Monte Carlo methods.}} Our focus here is simply on the nature
of the problem that was at the heart of \cite{metropolis:1953}, the solution
proposed, and the ultimate importance of that solution to Bayesian
computation.

In short, the authors wished to compute an expectation of the form, 
\begin{equation}
\mathbb{\mathbb{E}}(g(\mathbf{x))=}\int_{\mathcal{X}}g(\mathbf{x)}p(\mathbf{x%
})d\mathbf{x,}  \label{metrop}
\end{equation}%
where $p(\mathbf{x})$ denotes the so-called Boltzmann distribution of a set, 
$\mathbf{x,}$ of $N$ particles on $\mathbb{R}^{2}$. (See %
\citealp{robert:casella:2011}, Section 2.1, for all details.) Two particular
characteristics of (\ref{metrop}) are relevant to us here: \textit{i)} the
integral is of very high dimension, $2N$, with $N$ large; and \textit{ii)} $%
p(\mathbf{x})$ is generally known only up to its integrating constant. The
implication of \textit{i)} is that a basic rectangular integration method,
based on $L$ grid-points in each of the $2N$ directions, is infeasible,
having a computational burden of $L^{2N}$ or, equivalently, an approximation
error of $O(L^{-1/2N})$ (\citealp{kloek1978bayesian}). The implication of 
\textit{ii)} is that a Monte Carlo\textit{\ }(MC) estimate of (\ref{metrop}%
), based on $M$ $i.i.d.$ direct draws from $p(\mathbf{x})$, $\mathbf{x}%
^{(i)} $, $i=1,2,...,M$: $\widehat{E}_{MC}(g(\mathbf{x))}=\frac{1}{M}%
\sum\nolimits_{i=1}^{M}g(\mathbf{x}^{(i)}\mathbf{),}$ with approximation
error of $O(M^{-1/2})$ independent of dimension, is not available.\footnote{%
The authors actually make mention of a \textit{na\"ive} Monte Carlo method,
based on \textit{Uniform} sampling over the $2N$ dimensional space, followed
by a reweighting of the Uniform draws by a kernel of $p(\mathbf{x}).$ The
idea is dismissed, however, as `not practical'. In modern parlance, whilst
this method would yield an $O(M^{-1/2})$ approximation error, the constant
term within the order would be large, since the Uniform distribution used to
produce draws of $\mathbf{x} $ differs substantially from the \textit{actual}
distribution of $\mathbf{x}$, $p(\mathbf{x}).$}

Features \textit{i)} and \textit{ii)} --- either individually or in tandem
--- broadly characterize the posterior expectations in (\ref{gen_expect})
that are the focus of this review. Hence the relevance to Bayesian
computation of the solution offered by \cite{metropolis:1953} to the
non-Bayesian problem in (\ref{metrop}); and we describe their solution with
direct reference to (\ref{gen_expect}) {and the notation used therein}.

Specifically, the authors advocate computing an integral such as (\ref%
{gen_expect}) via the simulation of a \textit{Markov chain: }$\boldsymbol{%
\theta }^{(i)}$, $i=1,2,...,M$, with \textit{invariant distribution} $p(%
\boldsymbol{\theta }\mathbf{|y})$. The draw at iteration $i+1$ in the chain
is created by taking the value at the $ith$ iteration, $\boldsymbol{\theta }%
^{(i)}$, and perturbing it according to a random walk:\textit{\ }$%
\boldsymbol{\theta }^{c}=\boldsymbol{\theta }^{(i)}+\delta \boldsymbol{%
\varepsilon }$, where each element of $\boldsymbol{\varepsilon }$ is drawn
independently from $U(-1,1)$, and $\delta $ `tunes' the algorithm.\footnote{%
\cite{metropolis:1953} actually implemented their algorithm one element of $%
\boldsymbol{\theta }$ at a time, as a harbinger of the Gibbs sampler to
come. See \cite{robert:casella:2011} for more details.} The `candidate' draw 
$\boldsymbol{\theta }^{c}$ is accepted as draw $\boldsymbol{\theta }^{(i+1)}$
with probability: 
\begin{equation}
\alpha =\min \{p^{\ast }(\boldsymbol{\theta }^{c}\mathbf{|y})/p^{\ast }(%
\boldsymbol{\theta }^{(i)}\mathbf{|y}),1\},  \label{MH_RW_ratio}
\end{equation}%
where $p^{\ast }$ is a kernel of $p.$ Using the theory of reversible Markov
chains, it can be shown (see, for example, \citealp{tierney94}) that use of (%
\ref{MH_RW_ratio}) to determine the $(i+1)th$ value in the chain does indeed
produce a dependent sequence of draws with invariant distribution $p(%
\boldsymbol{\theta }\mathbf{|y})$. Hence, subject to convergence to $p(%
\boldsymbol{\theta }\mathbf{|y})$ (conditions for which were verified by the
authors\ for their particular problem) these draws can be be used to
estimate (\ref{gen_expect}) as the sample mean, 
\begin{equation}
\overline{g(\boldsymbol{\theta }\mathbf{)}}=\frac{1}{M}\sum%
\limits_{i=1}^{M}g(\boldsymbol{\theta }^{(i)}\mathbf{),}  \label{metrop_est}
\end{equation}%
and an appropriate weak law of large numbers and central limit
theorem invoked to prove the $\sqrt{M}$-consistency and limiting
normality of the {estimator}. (See \citealp{geyer2011introduction}, for
details.)

Due to the (positive) autocorrelation in the Markov chain, the variance of
the \textit{Metropolis estimator} (as it would become known) is larger than
that of the (infeasible) MC estimate in (\ref{MH_RW_ratio}), computed as in (%
\ref{metrop_est}), but using $i.i.\dot{d}$ draws from $p(\boldsymbol{\theta }%
\mathbf{|y})$, namely:%
\begin{equation}
\sigma _{MC}^{2}=\text{Var}(g(\boldsymbol{\theta }))/M,  \label{MC_variance}
\end{equation}%
{expressed here for the case of scalar }$g(\boldsymbol{\theta })$. However,
as is clear from (\ref{MH_RW_ratio}), the Metropolis MCMC algorithm requires
knowledge of $p(\boldsymbol{\theta }\mathbf{|y})$ only up to the normalizing
constant, and does \textit{not }require direct simulation from $p(%
\boldsymbol{\theta }\mathbf{|y})$ itself. It is this\textit{\ }particular
feature that would lend the technique its great power in the decades to come.%
\footnote{\cite{dongarra2000guest} rank the Metropolis algorithm as one of
the 10 algorithms \textquotedblleft with the greatest influence on the
development and practice of science and engineering in the 20th
century\textquotedblright .}

\subsection{1964: Hammersley and Handscomb: Importance Sampling}\label{hh}

The obviation of the need to directly\textit{\ }sample from $p(\boldsymbol{%
\theta }\mathbf{|y})$ also characterizes importance sampling, and underlies
its eventual importance in solving difficult Bayesian computational
problems. Nevertheless, \cite{hammersley:handscomb:1964} did not emphasize
this feature but, rather, focussed on the ability of IS to produce variance
reduction in simulation-based estimation of integrals.\footnote{%
One could in fact argue that a similar aim motivated Metropolis and
co-authors, given that they drew a sharp contrast (in effect) between the
efficiency of their method and that of the na\"ive Monte Carlo technique
based on Uniform sampling.} Again, the focus was not on Bayesian integrals,
but we describe the method in that setting.\footnote{%
Whilst \cite{hammersley:handscomb:1964} provides a textbook exposition of
importance sampling, the concept actually appeared in published form much
earlier, initially under the name of `quota sampling'. See \cite{kahn:1949}
and \cite{goert:kahn:1950} for example, plus \cite{andral:hal-03696061} for
a recent exploration into the historical origins of importance sampling.}

In brief, given an `importance' ({or }`proposal') density, $q(\boldsymbol{%
\theta }\mathbf{|y})$, that preferably mimics $p(\boldsymbol{\theta }\mathbf{%
|y})$ well, and $M$ $i.i.d.$ draws, $\boldsymbol{\theta }^{(i)}$, from $q(%
\boldsymbol{\theta }\mathbf{|y})$, an IS estimate of (\ref{gen_expect}) is $%
\overline{g(\boldsymbol{\theta }\mathbf{)}}_{IS}=\frac{1}{M}%
\sum\limits_{i=1}^{M}g(\boldsymbol{\theta }^{(i)}\mathbf{)}w(\boldsymbol{%
\theta }^{(i)})\mathbf{,}$ where $w(\boldsymbol{\theta }^{(i)})=p(%
\boldsymbol{\theta }^{(i)}\mathbf{|y})/q(\boldsymbol{\theta }^{(i)}\mathbf{|y%
}).$ In the typical case where $p(\boldsymbol{\theta }^{(i)}\mathbf{|y})$ is
available only up to the integrating constant, and $w(\boldsymbol{\theta }%
^{(i)})$ cannot be evaluated as a consequence, the estimate is modified as 
\begin{equation}
\overline{g(\boldsymbol{\theta }\mathbf{)}}_{IS}=\sum\limits_{i=1}^{M}g(%
\boldsymbol{\theta }^{(i)}\mathbf{)}w(\boldsymbol{\theta }^{(i)})\big/%
\sum_{j=1}^{M}w(\boldsymbol{\theta }^{(i)}),  \label{is_est_2}
\end{equation}%
with the weights re-defined as $w(\boldsymbol{\theta }^{(j)})=p^{\ast }(%
\boldsymbol{\theta }^{(i)}\mathbf{|y})/q^{\ast }(\boldsymbol{\theta }^{(i)}%
\mathbf{|y})$, for kernels, $p^{\ast }(\boldsymbol{\theta }^{(i)}\mathbf{|y}%
) $ and $q^{\ast }(\boldsymbol{\theta }^{(i)}\mathbf{|y})$, of $p(%
\boldsymbol{\theta }\mathbf{|y})$ and $q(\boldsymbol{\theta }\mathbf{|y})$
respectively. Once again, and under regularity conditions pertaining to the 
\textit{importance density }$q(\boldsymbol{\theta }\mathbf{|y)}$, asymptotic
theory can be invoked to prove that (\ref{is_est_2}) is a $\sqrt{M}$%
-consistent estimator of $\mathbb{\mathbb{E}}(g(\boldsymbol{\theta }\mathbf{%
)|y)}$ \citep{geweke:1989}. A judicious choice of $q(\boldsymbol{\theta }%
\mathbf{|y)}$ is able to yield a sampling variance that is less than (\ref%
{MC_variance}) in some cases, as befits the original motivation of IS as a
variance reduction method. {(See \citealp{geweke:1989}, and %
\citealp{robert:casella:2004}, for discussion.}) Critically however, like
the Metropolis method, (\ref{is_est_2}) serves as a feasible estimate of $%
\mathbb{\mathbb{E}}(g(\boldsymbol{\theta }\mathbf{)|y)}$ when $p(\boldsymbol{%
\theta }\mathbf{|y})$ cannot be easily simulated; hence the significance of
IS in Bayesian computation. Moreover, its maintenance of independent\textit{%
\ }draws, allied with its re-weighting of draws from an approximating
density, has led to the emergence of IS as a vehicle for implementing\ SMC
algorithms, to be referenced in Section \ref{exact}.

\subsection{1970: Hastings and his Generalization of the Metropolis
Algorithm}\label{hast copy(1)}

The final publication that we pinpoint during the 200-odd year period
subsequent to 1763, is the 1970 \textit{Biometrika }paper, `\textit{Monte
Carlo Sampling Methods Using Markov Chains and Their Applications', }by
Wilfred Keith Hastings. Whilst \cite{metropolis:1953} proposed the use of
MCMC sampling to compute particular integrals in statistical mechanics, it
was the Hastings paper that elevated the concept to a general one, and
introduced it to the broader statistics community. Included in the paper is
also the first mention of what would become known as the \textit{%
Metropolis-within-Gibbs sampler} \citep{robert:casella:2011}. Once again,
the author's focus was not a Bayesian integral \textit{per se}; however we
describe the method in that context.

In contrast to Metropolis and co-authors, \cite{hastings:1970} acknowledges
up-front that the need to know $p(\boldsymbol{\theta }\mathbf{|y})$ only up
to the integrating constant is a compelling feature of an MCMC-based
estimate of (\ref{gen_expect}). Hastings also generalizes the acceptance
probability in (\ref{MH_RW_ratio}) to one that accommodates a general
`candidate' distribution $q(\boldsymbol{\theta }\mathbf{|y)}$ from which $%
\boldsymbol{\theta }^{c}$ is drawn, as:%
\begin{equation}
\alpha = \dfrac{p^{\ast }(\boldsymbol{\theta }^{c}\mathbf{|y})/q(
\boldsymbol{\theta }^{(i)}|\boldsymbol{\theta }^{c},\mathbf{y)}}{p^{\ast }(\boldsymbol{\theta }^{(i)}\mathbf{|y})/q(\boldsymbol{\theta 
}^{c}|\boldsymbol{\theta }^{(i)},\mathbf{y)}}\wedge 1
\label{hastings}
\end{equation}%
which clearly collapses to (\ref{MH_RW_ratio}) when $q(\boldsymbol{\theta }%
\mathbf{|y)}$ is symmetric (in $\boldsymbol{\theta }^{c}$ and $\boldsymbol{%
\theta }^{(i)}$), as in the original random walk proposal of \cite%
{metropolis:1953}. Importantly, the more general algorithm allows for a
targeted choice of $q(\boldsymbol{\theta }\mathbf{|y)}$ that reduces the
need for tuning and which can, potentially, reduce the degree of dependence
in the chain and, hence, the variance of the estimate of $\mathbb{E}(g(%
\boldsymbol{\theta }\mathbf{)|y).}$ Hastings formalizes the standard error
of this estimate using time series theory, explicitly linking, for the first
time, the autocorrelation in the Markov draws to the efficiency of the
MCMC-based estimate of (\ref{gen_expect}). Crucially, the author tackles the
issue of dimension by advocating the treatment of one element (or several
elements) of $\boldsymbol{\theta }$ at a time, conditional on all remaining
elements.

In summary, all of the important ingredients from which the huge smorgasbord
of future MCMC algorithms would eventually be constructed --- for the 
\textit{express} purpose of solving Bayesian problems --- were now on the
table, via this particular paper.

\section{The Late 20th Century: Gibbs Sampling \& the MCMC Revolution}
\label{70s}

\begin{figure}[th]
\centering
\includegraphics[scale=0.6]{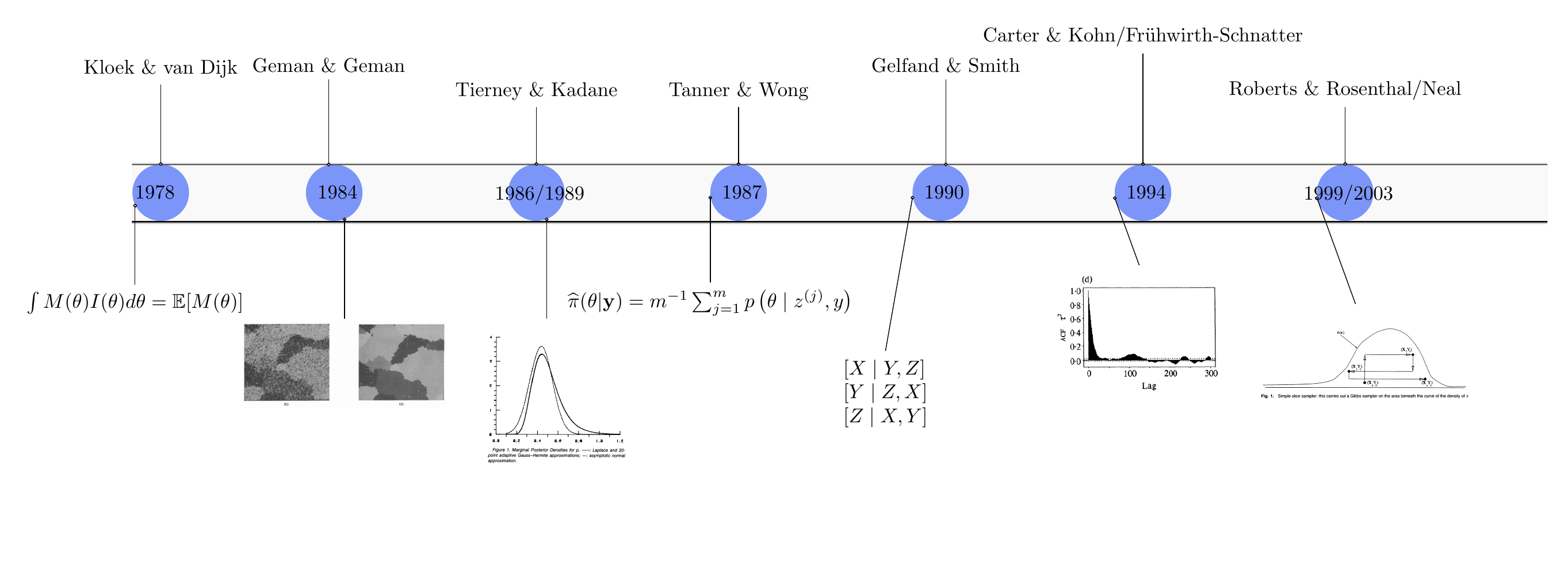}
\caption{From left to right: \textit{1)} Equation (3.3) in \protect\cite%
{kloek1978bayesian}, where $M(\boldsymbol{\protect\theta })=g(\boldsymbol{%
\protect\theta })\protect\kappa (\boldsymbol{\protect\theta }|\mathbf{Y},%
\mathbf{Z})p(\boldsymbol{\protect\theta })/I(\boldsymbol{\protect\theta })$,
with $\protect\kappa (\boldsymbol{\protect\theta }|\mathbf{Y},\mathbf{Z})$
the likelihood function, $\mathbf{Y}$ and $\mathbf{Z}$ matrices of
observations on endogenous and predetermined variables respectively, $p(%
\boldsymbol{\protect\theta })$ the prior, $g(\boldsymbol{\protect\theta })$
defining the parameter function of interest, and $I(\boldsymbol{\protect%
\theta })$ the importance density. The first application of importance
sampling to compute a posterior expectation; \textit{2)} Figure 2 in 
\protect\cite{geman:1984}; Gibbs sampling is used to restore image (d) from
the degraded image (b); \textit{3)} Figure 1 in \protect\cite%
{tierney:kadane:1986}, illustrating the accuracy of a Laplace approximation
of the marginal posterior for the parameter $\protect\rho $; \textit{4)} The
exploitation of `data augmentation' to estimate the marginal posterior of $%
\boldsymbol{\protect\theta }$ as an average of conditional posteriors given $%
m$ draws of the latent state vector, $\mathbf{z}$; Page 530 in \protect\cite%
{tanner87}; \textit{5)} The sequence of conditional distributions that
define the Gibbs sampler in \protect\cite{gelfand:smith90}; \textit{6)} An
illustration from Figure 2 in \protect\cite{carter:kohn:1994} of the rapid
decline in the sample autocorrelation function for draws from their
multi-state Gibbs sampling algorithm; proposed independently by \protect\cite%
{fruhwirth-schnatter:1994} as the forward-filtering-backward-sampling
(`FFBS') algorithm; \textit{7)} A diagrammatic representation of the slice
sampler in Figure 1 of \protect\cite{roberts:rosenthal:1999}.}
\end{figure}

Whilst the role that \textit{could }be played by simulation in computation
was thus known by the 1970s, the computing technology needed to exploit that
knowledge lagged behind.\footnote{%
Many readers may be too young to remember the punchcards! But there was a
time when RAND's 1955 \textit{A Million Random Digits with 100,000 Normal
Deviates} was more than an entry for sarcastic Amazon comments, as producing
this million digits took more than two months at the time.} Over the next
two decades, however, things changed. Indeed, \textit{two} developments now
went hand in hand to spawn a remarkable expansion in simulation-based
Bayesian computation:\textit{\ i)} the increased speed and availability of
computers, including personal desktop computers (\citealp{ceruzzi:2003}),
and \textit{ii)} the collective recognition that MCMC draws from a joint
posterior, $p(\boldsymbol{\theta }\mathbf{|y})$, could be produced via
iterative sampling from lower dimensional, and often standard, \textit{%
conditional }posteriors. When allied with both the concept of \textit{%
augmentation, }and an understanding of the theoretical properties of
combinations of MCMC algorithms, \textit{ii) }would lead to \textit{Gibbs
sampling} (with or without \textit{Metropolis-Hastings }(MH) subchains)
becoming the work-horse of Bayesian computation in the 1990s.

An MH algorithm `works', in the sense of producing a Markov chain that
converges to the required distribution $p(\boldsymbol{\theta }\mathbf{|y})$,
due to the form of the acceptance probability in (\ref{hastings}) (or the
nested version in (\ref{MH_RW_ratio})). More formally, the algorithm, as
based on candidate density $q(\boldsymbol{\theta }\mathbf{|y)}$, and
acceptance probability as defined in (\ref{hastings}), defines a \textit{%
transition kernel }with invariant distribution, $p(\boldsymbol{\theta }%
\mathbf{|y}).$ The `Gibbs sampler' similarly yields a Markov chain with
invariant distribution, $p(\boldsymbol{\theta }\mathbf{|y})$, but via a
transition kernel that is defined as the product of full conditional
posteriors\textit{\ }associated with the joint. For the simplest case of a
two-dimensional vector $\boldsymbol{\theta }=(\theta _{1},\theta
_{2})^{\prime }$, the steps of the Gibbs algorithm are as follows: first,
specify an initial value for $\theta _{2}$, $\theta _{2}^{(0)}$; second, for 
$i=1,2,...,M$, cycle iteratively through the two conditional distributions,
drawing respectively: $\theta _{1}^{(i)}$ from $p_{1}(\theta
_{1}^{(i)}|\theta _{2}^{(i-1)},\mathbf{y})$, and $\theta _{2}^{(i)}$ from $%
p_{2}(\theta _{2}^{(i)}|\theta _{1}^{(i)},\mathbf{y)}$. Given the
satisfaction of the required convergence conditions (which essentially place
sufficient regularity on the conditionals), the draws $\boldsymbol{\theta }%
^{(i)}=(\theta _{1}^{(i)},\theta _{2}^{(i)})^{\prime }$, $i=1,2,...,M$,
converge in distribution to the joint posterior distribution as $%
M\rightarrow \infty $, and can be used to produce a $\sqrt{M}$-consistent
estimator of (\ref{gen_expect}) in the form of (\ref{metrop_est}). Extension
to higher-dimensional problems is obvious, although decisions about how to
`block' the parameters, and thereby define the conditionals, now play a role %
\citep{roberts:sahu:1997}.\footnote{%
The Gibbs sampler can be viewed as (and in some expositions is presented as)
a special case of a `multiple-block' MH sampler, in which the candidate
values for each block of parameters are drawn directly from their full
conditional distributions and the acceptance probability in (each blocked
version of) (\ref{hastings}) is equal to one. (See, for example, Chib,
2011). See also \cite{tran2018common} for further discussion of this point
--- conducted in the context of a \textit{generalized} MH framework, in which
many of the MCMC algorithms mentioned in Sections \ref{advances} are also
nested.}

Gibbs thus exploits the simplicity yielded by conditioning: whilst joint and
marginal posterior distributions are usually complex in form, (full)
conditional posteriors are often standard and, hence, able to be simulated
from directly. While one may find hints in both \cite{hastings:1970} and 
\cite{besag:1974}, this point was first made clearly by \cite{geman:1984},
who also coined the phrase `Gibbs sampling' because their problem used Gibbs
random fields in image restoration (named, in turn, after the physicist,
Josiah Willard Gibbs). However, the later paper by \cite{gelfand:smith90} is
generally credited with bringing this transformational idea to the attention
of the statistical community, and illustrating its broad
applicability.

The idea of Gibbs sampling overlapped with a related proposal by \cite%
{tanner87}: that of `augmenting' the set of unknowns ($\boldsymbol{\theta }$
in our notation) with latent data, $\mathbf{x}=(x_{1},x_{2},...,x_{n})^{%
\prime }$, to yield conditionals --- $p(\boldsymbol{\theta }|\mathbf{x},%
\mathbf{y})$ and $p(\mathbf{x}|\boldsymbol{\theta },\mathbf{y})$ --- that
facilitate the production of a simulation-based estimate of $p(\boldsymbol{%
\theta }\mathbf{|y})$; with $p(\boldsymbol{\theta }|\mathbf{x},\mathbf{y})$,
in particular, often being standard. The melding of these two ideas, i.e.
sampling via conditionals \textit{per se}, and yielding more tractable
conditionals through the process of augmentation, enabled the analysis of
complex models that had thus far eluded Bayesian treatment, due to their
dependence on high-dimensional vectors of latent variables; selected
examples being: \cite{carlin:polson:stoffer:1992}, \cite{carter:kohn:1994}, 
\cite{fruhwirth-schnatter:1994} and \cite{jacquier94}.\ However, it also led
to the realization that \textit{auxiliary }latent variables could be
judiciously introduced into a model for the sole purpose of producing
tractable conditional posteriors over the augmented space, thereby opening
up a whole range of additional models to a Gibbs-based solution (e.g. %
\citealp{albert:chib:1993b}; \citealp{dieb:robe:1994}; {\citealp{higdon1998};%
} \citealp{kim1998svl}; \citealp{damien:wakefield:walker:1999}). {The 
\textit{slice sampler} (\citealp{roberts:rosenthal:1999}; \citealp{neal:2003}%
) is one particularly notable, and generic, way of generating an MCMC
algorithm via this principle of auxiliary variable augmentation.}

Of course, in most high-dimensional problems --- and in particular those in
which latent variables feature --- certain conditionals remain nonstandard,
such that direct simulation from them is not possible. Critically though,
the reduced dimension renders this a simpler problem than sampling from the
joint itself: via either the inverse cumulative distribution function
technique \citep{devroye:1986} --- approximated in the `Griddy Gibbs'
algorithm\textbf{\ }of \cite{ritter:tanner:1992} --- or by embedding an MH
algorithm within the outer Gibbs loop (a so-called `Metropolis-within-Gibbs'
algorithm).\footnote{%
We refer the reader to: \cite{besag:green:1993}, \cite{smith:roberts:1993}
and \cite{chib_greenberg_1996} for early reviews of MCMC sampling; \cite%
{casella:george:1992} and \cite{chibandgreenberg:1995} for descriptions of
the Gibbs and MH algorithms (respectively) that are useful for
practitioners; \cite{robert2015metropolishastings}, \cite{betancourt:2018}
and \cite{dunson2019hastings} for more recent reviews; and \cite%
{andrieu2004computational} and \cite{robert:casella:2011} for historical
accounts of MCMC sampling.}

\section{The 21st Century: A Second Computational Revolution!}\label{second}

\begin{figure}[th]
\centering
\includegraphics[scale=0.6]{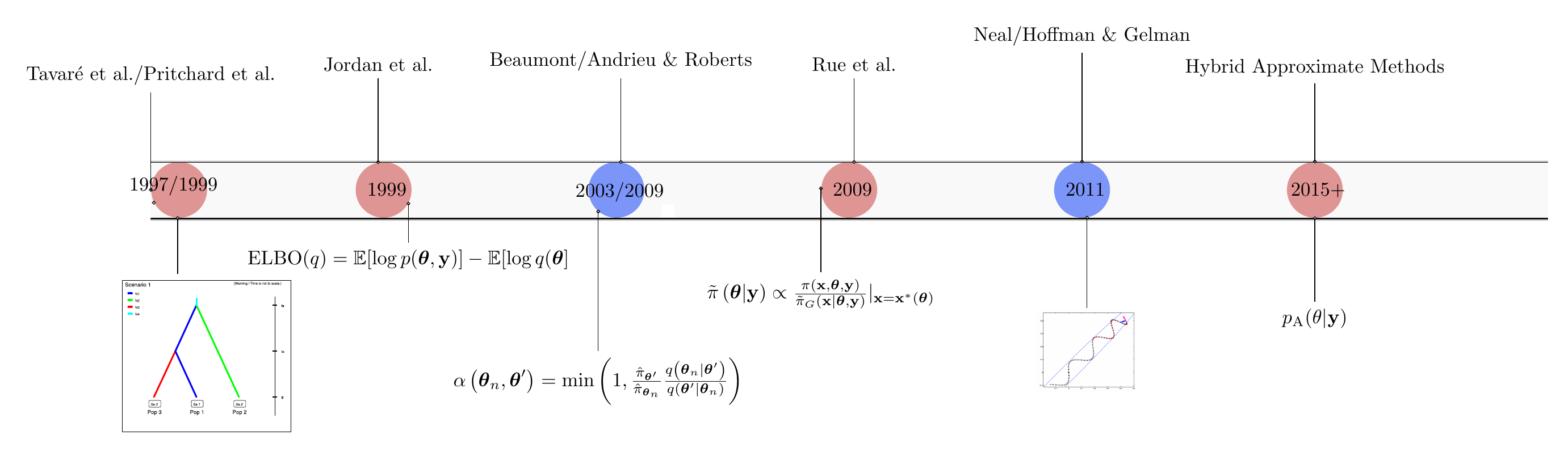}
\caption{From left to right: \textit{1)} An example of a phylogenic tree,
which represents one of the earliest types of model to which approximate
Bayesian computation (ABC) was applied; \protect\cite%
{tavare:balding:griffith:donnelly:1997} and \protect\cite%
{pritchard:seielstad:perez:feldman:1999} were the first to apply the method
of ABC to intractable models in population genetics; \textit{2) }The
`evidence lower bound' (ELBO), maximized with respect to the variational
density, $q(\boldsymbol{\protect\theta })$, in order to produce a
variational Bayes approximation to the posterior $p(\boldsymbol{\protect%
\theta }|\mathbf{y}).$ \protect\cite{jordanetal:1999} were the first to
apply the method to high-dimensional graphical models; \textit{3) }The
acceptance probability used in a pseudo-marginal Metropolis Hastings
algorithm, as based on an estimate of the likelihood function defining the
kernel of the posterior, $\protect\pi $, and a proposal distribution $q$; 
\textit{4) }The approximation of the posterior $\protect\pi (\boldsymbol{%
\protect\theta }|\mathbf{y})$ based on integrated nested Laplace
approximation (INLA), with $\protect\widetilde{\protect\pi }_{G}(\mathbf{x}|%
\boldsymbol{\protect\theta },\mathbf{y})$ a Gaussian approximation of $%
\protect\pi (\mathbf{x}|\boldsymbol{\protect\theta },\mathbf{y})$ and $%
\mathbf{x}^{\ast }\mathbf{(\boldsymbol{\protect\theta })}$ the mode of $p(%
\mathbf{x},\boldsymbol{\protect\theta },\mathbf{y})$ (for a given $%
\boldsymbol{\protect\theta }$); Equation (3) in the seminal INLA paper of 
\protect\cite{rue:martino:chopin:2009}; \textit{5) }A trajectory generated
during one iteration of the No-U-Turn Sampler; an extension of Hamiltonian
Monte Carlo; Figure 2 in \protect\cite{hoffman2014no}; \textit{6) }The
myriad of methods for producing an approximate posterior, $p_{A}(\boldsymbol{%
\protect\theta }|\mathbf{y})$, that are `hybrids' of various individual
approximate methods, are reviewed in Section 3.3 of \protect\cite%
{martin2021approximating}.}
\end{figure}

All of the methodological and theoretical developments referenced above,
allied with advances in computer hardware and software, spawned a huge
flowering of simulation-based Bayesian computation \textbf{---} across all
disciplines \textbf{---} in the final decades of the 20th century, with
variants of MCMC leading the charge. (See, for example, %
\citealp{brooks:etal:2011}, and \citealp{geweke2011handbook}, for extensive
coverages of the empirical problems to which MCMC algorithms (in particular)
have been\textbf{\ }applied.) However, the early algorithms did stumble in
the case of so-called `intractable' Bayesian problems, namely: \textit{1)} a
DGP that cannot readily be expressed as a probability density or mass
function (the `unavailable likelihood' problem); \textit{2)} a very large
dimension for $\boldsymbol{\theta }$ (the `high-dimensional' problem);
and/or \textit{3)} a very large dimension for $\mathbf{y}$ (the `big-data'
problem). See \cite{rue:martino:chopin:2009}, \cite{greenetal2015}, \cite%
{bardenet2017markov}, \cite{blei2017variational}, \cite{betancourt:2018}, 
\cite{robert2018accelerating}, \cite{johndrow2019mcmc}, \cite{Jahan2020},
and \cite{martin2021approximating} for relevant discussions.

In response, a wealth of solutions have been proposed. For the purpose of
this paper it is convenient to categorize these solutions as either `exact'
or `approximate'. Exact solutions still invoke MCMC or IS principles to
compute (\ref{gen_expect}) in cases where one or more form of intractability
obtains. That is, the goal of such methods is to still estimate the
posterior expectation in (\ref{gen_expect}) `exactly', at least up to an
order $O(M^{-1/2})$, where $M$ is the number of draws that defines the
simulation scheme, and which can --- in principle --- be made arbitrarily
large.\footnote{%
We note here so-called `quasi-Monte Carlo' integration schemes, which aim
for exactness at a faster rate than $O(M^{-1/2})$. See \cite%
{lemieux2009monte} for a review of such methods, \cite{chen2011} for the
extension to quasi-MCMC algorithms, and \cite{gerber:chopin:2015} for an
entry on sequential quasi-Monte Carlo.} Approximate solutions, on the other
hand, use computation to target only an \textit{approximation} --- of one
sort or another --- of the expectation in (\ref{gen_expect}).

\subsection{Exact Solutions to Intractable Bayesian Problems}\label{exact}

\subsubsection{Advances in MCMC.}\label{advances}

We begin with four reminders about MCMC algorithms:

\begin{enumerate}
\item \textit{First: }MCMC methods avoid the need to simulate from a joint
posterior directly by simulating the unknowns indirectly, via another
distribution (or set of distributions) from which simulation is feasible.
However, such methods still require the evaluation of the DGP as a
probability density function or a probability mass function: either in the
computation of the acceptance probability in (\ref{MH_RW_ratio}) or (\ref%
{hastings}) in a Metropolis-type algorithm, or in the implementation of any
Gibbs-based algorithm, in which the conditional posteriors are required
either in full form or at least up to a scale factor.

\item \textit{Second}: an \textit{MC}MC algorithm is just that --- a 
\textit{Markov chain }Monte Carlo algorithm. As such, an MCMC scheme --- by
design --- produces a \textit{local }exploration of the target posterior,
with the location in the parameter space of any simulated draw being
dependent\textit{\ }on the location of the previous draw, in a manner that
reflects the specific structure of the algorithm. Most notably, an MCMC
algorithm with a high degree of dependence will potentially be slow in
exploring the high mass region of the target posterior (or the `target set',
in the language of \citealp{betancourt:2018}), with this problem usually
being more severe the larger is the dimension of the parameter space. Looked
at through another lens: for $M$ MCMC draws, the greater the degree of
(typically positive) dependence in those draws, the less efficient is the
MCMC-based estimate of (\ref{gen_expect}), relative to an estimate based on $%
M$ $i.i.d.$ draws from the target. This loss of efficiency is measured by
the so-called \textit{inefficiency factor }(IF), defined ({in the case of
scalar }$g(\boldsymbol{\theta })$) as the ratio of the \textit{MCMC standard
error}, $\sigma _{MCMC}$ with $\sigma _{MCMC}^{2}={\text{Var}(g(%
\boldsymbol{\theta }))[1+2\sum\nolimits_{l=1}^{\infty }\rho _{l}]/M}$ to the
standard error associated with $M$ $i.i.d.$ draws, $\sigma _{MC}^{2}$, with $%
\sigma _{MC}^{2}$ as given in (\ref{MC_variance}), where $\rho _{l}$ is the
lag-$l$ autocorrelation of the draws of $g(\boldsymbol{\theta })$ over the
history of the chain. This ratio, in turn, defines the \textit{effective
sample size }of the MCMC algorithm, $ESS=M/[1+2\sum\nolimits_{l=1}^{\infty
}\rho _{l}].$ Improving the efficiency of an MCMC algorithm, for any given
value of $M$, thus equates to increasing $ESS$ to its maximum possible value
of $M$ by reducing the dependence in the draws.

\item \textit{Third}: an MC\textit{MC} algorithm is also a Markov chain%
\textit{\ Monte Carlo} algorithm. That is, under appropriate regularity it
produces a $\sqrt{M}$-consistent estimate of (\ref{gen_expect}), whatever
the degree of dependence in the chain, with the dependence affecting the
constant term implicit in the $O(M^{-1/2})$ rate of convergence, but not the
rate itself. Hence, in principle, any MCMC algorithm, no matter how
inherently inefficient, can produce an estimate of (\ref{gen_expect}) that
is arbitrarily accurate, simply through an increase in $M.$ However, an
increase in $M$ entails an increase in \textit{computational cost},
measured, say, by\textit{\ computing clock-time}. The extent of this
increase depends, in turn, on the (per-iteration) cost of generating a
(proposal/candidate) draw and, with an MH step, the cost of calculating the
acceptance probability. Both component costs will (for any algorithm)
clearly increase with the number of unknowns that need to be simulated, and
assessed, at each iteration. Either cost, or both, will also increase with
the sample size, given the need for pointwise evaluation of the likelihood
function across the elements of $\mathbf{y}$.

\item \textit{Fourth}: the very concept of efficiency is relevant only if
the Markov chain is (asymptotically in $M$) \textit{unbiased}, which depends
critically on draws being produced from the correct invariant distribution.
That is, the production of an accurate\textit{\ }MCMC-based estimate of (\ref%
{gen_expect}) depends, not just on reducing the degree of dependence in the
chain, or on increasing the number of draws, but on ensuring that the chain%
\textit{\ actually} explores the target set, and thereby avoids bias in the
estimation of (\ref{gen_expect}).\footnote{%
It is acknowledged in the literature that MCMC algorithms produce
potentially strong biases in their initial phase of `convergence' to the
typical set from an initial point in the parameter space. However, under
appropriate regularity, such biases are transient, and their impact on the
estimation of (\ref{gen_expect}) able to be eliminated by discarding a
sufficiently large number of `burn-in' or `warm-up' draws from the
computation. (See \citealp{robert:casella:2004}, and %
\citealp{gelman2011inference}, for textbook discussions of convergence,
including diagnostic methods.) Some of the more recent literature is
concerned with removing this transitory bias after a finite number of
iterations; e.g. \cite{jacob2019unbiased}. Other literature is concerned
with ensuring that an MCMC algorithm does not yield a bias that is \textit{%
non-transitory} due to the inability of the algorithm to effectively explore
the target set at all (within a meaningful time frame); see e.g. \cite%
{betancourt:2018}.}
\end{enumerate}

With reference to Point 1: so-called \textit{pseudo-marginal }MCMC methods
can be used to obviate the problem of the DGP either being unavailable or
being computationally challenging, by inserting within an MCMC algorithm an 
\textit{unbiased} estimate of $p(\mathbf{y}|\boldsymbol{\theta })$, thereby
retaining $p(\boldsymbol{\theta }\mathbf{|y})$ as the invariant distribution
of the chain (\citealp{beaumont:2003}; \citealp{andrieu:roberts:2009}). The
use of an estimate of the likelihood in an MH algorithm prompted use of the
term `pseudo-marginal' MH (PMMH) by \cite{andrieu:roberts:2009} although, as
noted, this replacement still yields a chain with an invariant distribution
equal to $p(\boldsymbol{\theta }\mathbf{|y})$ when
the estimate is unbiased; hence the method remains `exact'. When a
likelihood estimate is produced specifically via the use of \textit{particle
filtering} in a state space model (SSM), the term \textit{particle} MCMC has
also been coined \citep{andrieu:doucet:holenstein:2010}.\footnote{%
Whilst not a pseudo-marginal method, particle filtering has also been used
to provide an estimate of $p(\mathbf{x}|\boldsymbol{\theta ,y})$ in a Gibbs
scheme for an SSM --- so-called `particle Gibbs' %
\citep{andrieu:doucet:holenstein:2010}.}

Whilst unbiasedness of the likelihood estimate is required for a general
PMMH algorithm to `work', the variance of the estimate also affects the
performance of the sampler and, hence, the simulation efficiency of any
estimate of (\ref{gen_expect}) that is produced. However, improving the
precision of the likelihood estimator comes at a computational cost, and an
`optimal' number of draws that balances computational cost with an
acceptable mixing of the chain needs to be sought. See \cite{pitt2012some}, 
\cite{doucet2015efficient} and \cite{deligiannidis2018correlated} for
discussion of the optimal structuring and tuning of pseudo-marginal
algorithms.

With reference to Points 2 to 4: many other advances in MCMC (some of which
also exploit pseudo-marginal principles) aim to increase the effectiveness\
with which an algorithm explores the high mass region of the target
posterior and, hence, the accuracy with which (\ref{gen_expect}) is
estimated, by doing one (or more) of three things: reducing dependence in
the chain, reducing the computational cost per iteration of the chain (thus
enabling more draws to be produced), or eliminating bias. In particular,
with reference to the taxonomy of `intractable' problems given earlier,
focus is increasingly directed towards developing algorithms that \textit{%
scale well}, in terms of the dimension of the data and/or the number of
unknowns. With our goal of brevity in mind, we simply list the main
contenders here, including certain key references or reviews, deflecting
both to those papers, and to the broad overviews of modern developments in
MCMC in \cite{greenetal2015}, \cite{robert2018accelerating} and \cite%
{dunson2019hastings} for all details. Of particular note is the recent
survey on Bayesian methods for `Big Data' in \cite{Jahan2020} (Section 5.1
being most pertinent), which describes the precise manner in which certain
of the methods cited below (and others) tackle the problem of scale. We
categorize the methods according to whether improved performance is achieved
(primarily): \textit{i)} via the exploitation of more geometric information
about the target posterior; \textit{ii)} by better choice of {proposals}; 
\textit{iii)} by the use of parallel, batched, subsample, coupled or
ensemble sampling methods; or \textit{iv)} by {the explicit use of variance
reduction methods.}

\begin{enumerate}
\item[\textit{i)}] Hamiltonian Monte Carlo (HMC) (\citealp{neal:2011}; %
\citealp{carpenter:etal:2017}; \citealp{betancourt:2018}); no U-turn
sampling (NUTS) (\citealp{hoffman2014no});\footnote{%
As described in \cite{neal:2011}, simulation methods based on Hamiltonian
dynamics can actually be viewed as having as long a history as MCMC itself.
The more modern manifestations of HMC, however, including NUTS, can be
viewed as Markov chain algorithms that simply explore the parameter space
more effectively than (say) a default random walk scheme. The probabilistic
programming platform Stan (\citealp{carpenter:etal:2017}) enables
implementation of NUTS.} Metropolis-Adjusted Langevin algorithm (%
\citealp{roberts1996exponential}; \citealp{roberts1998optimal}); stochastic
gradient MCMC \citep{nemeth2019stochastic}; piecewise deterministic Markov
processes (\citealp{bierkens2018piecewise}; \citealp{fearnhead:etal:2018}; %
\citealp{bierkens2019}).

\item[\textit{ii)}] Optimal scaling of random-walk MH %
\citep{roberts:gelman:gilks:1997}; adaptive sampling (%
\citealp{nott2005adaptive}; \citealp{roberts2009examples}; %
\citealp{rosenthal2011optimal}); {MCMC with ordered overrelaxation (%
\citealp{Neal1998}); simulated tempering, parallel tempering and tempered
transition methods} (\citealp{geyer91pt}; \citealp{marinarietparisi92}; %
\citealp{Neal1996};\textbf{\ }\citealp{Gramacy2010}; \citealp{geyer2011st}; %
\citealp{tawn2019weight}); delayed rejection sampling (%
\citealp{tierney:mira:1998}); delayed acceptance sampling (%
\citealp{christen:fox:2005}; \citealp{Golightly2015}; %
\citealp{wiqvist2018accelerating}; \citealp{banterle2019accelerating});
multiple try MCMC (\citealp{liuliang2000}; \citealp{bedard2012scaling}; %
\citealp{martino2018review}; \citealp{luo2019multiple}); taylored randomized
block MH (\citealp{CHIB201019}); tempered Gibbs sampling (%
\citealp{zanellaroberts2019}); quasi-stationary Monte Carlo and subsampling %
\citep{pollock:etal:2020}.

\item[\textit{iii)}] Parallelized MCMC (\citealp{jacob:robert:smith:2010}; %
\citealp{wang:dunson:2013}); subposterior (batched) methods (%
\citealp{neiswanger:wang:xing:2013}; \citealp{scott2016bayes}); subsampling
methods based on pseudo-marginal MCMC (\citealp{bardenet2017markov}; %
\citealp{quiroz2018speeding}; \citealp{quiroz2019speeding}); {perfect
sampling (\citealp{proppetwilson96}; \citealp{casella:lavine:2001}, %
\citealp{craiu2011perfection}; \citealp{huber2016perfect}); }unbiased MCMC
via coupling (\citealp{glynn:rhee:2014}; \citealp{Glynn:exact2016}; %
\citealp{middleton2018unbiased}; \citealp{jacob2019unbiased}); unbiased MCMC
using pseudo-marginal principles (\citealp{lyne2015}); ensemble MCMC (%
\citealp{iba0}; \citealp{cappe:guillin:marin:robert:2004}; %
\citealp{neal2011mcmc}).

\item[\textit{iv)}] Rao-Blackwellization (\citealp{casella:robert:1996}; %
\citealp{robert:casella:2004}; \citealp{douc:robert:2011}); {antithetic
variables (\citealp{Frigessi2000}; \citealp{craiu:meng:2005}); control
variates (\citealp{Dellaportas2012}; \citealp{baker2019}); }thinning (%
\citealp{owen2017statistically}).
\end{enumerate}

\subsubsection{The role of SMC.}

Sequential Monte Carlo (SMC) methods \textbf{---} which involve the
sequential application of importance sampling steps \textbf{---} have also
evolved over the 21st century. Most notably, they have expanded from being
`particle filters' designed for the sequential analysis of state space
models, \citep{gordon:salmon:smith:1993}, into a broader set of techniques
used to perform sequential tasks in non-state space settings; we refer the
reader to \cite{naesseth2019elements} and \cite{Chopin2020} for extensive
reviews. From the perspective of dealing with intractable problems, their
role in particle MCMC (as described above) has been particularly critical,
and in playing that role they have retained their status as `exact' methods.
However, they have also played an important role in the context of certain
of the approximate methods to be discussed below, via: ABC-SMC algorithms (%
\citealp{sisson:fan:tanaka:2007}; %
\citealp{beaumont:cornuet:marin:robert:2009}), `ABC filtering' (%
\citealp{jasra:etal:2012}; \citealp{dean2014parameter}; %
\citealp{calvet2015accurate}; \citealp{jasra2015approximate}), and VB with
intractable likelihood (\citealp{tran2017variational}).

\subsection{Approximation Solutions to Intractable Bayesian Problems}
\label{approx}

In contrast to exact methods of computation, when applying an approximate
method investigators do not seek exactness, other than perhaps claiming
asymptotic (in $n$) validity under certain conditions. That is, for finite $%
n $ at least, such methods only ever provide a numerical solution to some 
\textit{approximation} of (\ref{gen_expect}). The advantage of such methods,
however, is that they can yield `reasonable' solutions, and often quickly,
to empirical problems that would test the limits of exact methods or,
indeed, be otherwise infeasible.

It is convenient, for the purpose of this review, to categorize the main
approximate methods according to whether their primary goal is to obviate
the need to evaluate the DGP (i.e. to solve the so-called \textit{%
doubly-intractable }problem), or to tackle problems of dimension (either in
the data, or the unknowns or, typically, in both). This categorization
corresponds to a distinction between \textit{simulation-based} approximate
methods and approximate methods that are primarily based on \textit{%
optimization}, and it is those categories that we use for the sub-section
headings below.

\subsubsection{Simulation-based methods: ABC and BSL.}

In computing (\ref{gen_expect}), both ABC and BSL replace the posterior in
the integrand, $p(\boldsymbol{\theta }\mathbf{|y})$, with a posterior that
conditions only on some \textit{function} of the observed data, $\mathbf{y.}$
Typically, that function is a small-dimensional vector of summary
statistics, $\eta (\mathbf{y})$, that are chosen to be `informative' about $%
\boldsymbol{\theta }$, in which case the posterior that is targeted is
represented as $p(\boldsymbol{\theta }\mathbf{|}\eta (\mathbf{y})).$ Using $%
M $ draws from $p(\boldsymbol{\theta }\mathbf{|}\eta (\mathbf{y}))$, $%
\boldsymbol{\theta }^{(i)}$, the sample mean, $\overline{g(\boldsymbol{%
\theta }\mathbf{)}}=(1/M)\sum\nolimits_{i=1}^{M}g(\boldsymbol{\theta }^{(i)}%
\mathbf{)}$, is used to estimate the expectation in (\ref{gen_expect}). The
key here is that producing draws from $p(\boldsymbol{\theta }\mathbf{|}\eta (%
\mathbf{y}))$ \textbf{---} and the precise way in which this is undertaken
differs between the two methods \textbf{---} entails only simulation of the
DGP, \textit{not }its evaluation; hence the suitability of ABC and BSL for
problems in which evaluation of the DGP is infeasible or, at the very least,
computationally challenging. What is lost of course, is quantification of
uncertainty about $\boldsymbol{\theta }$ that is informed by the full data
set; inference is only `partial', with the choice of summaries that are
`close' (in some sense) to being sufficient being an important goal. We
refer the reader to \cite{marin:pudlo:robert:ryder:2011}, \cite%
{price2018bayesian}, \cite{Beaumont2019} and \cite{sisson2018handbook}, for
details and extensive referencing.

\subsubsection{Optimization methods: VB and INLA.}

VB and INLA target alternative approximations to (\ref{gen_expect}). Both
methods are particularly beneficial when the dimension of the unknowns, or
the dimension of the data, or both, are very large, usually due to the
presence in the model of a high number of latent, or `local', parameters, $%
\mathbf{x}$, in addition to the smaller set of `global' (or
`hyperparameters) parameters, $\boldsymbol{\theta }$.

Using the principle of the calculus of variations, VB produces an
approximation of $p(\boldsymbol{\theta }\mathbf{|y})$, $q^{\ast }(%
\boldsymbol{\theta })$ say, that is `closest' to $p(\boldsymbol{\theta }%
\mathbf{|y})$ within a chosen family of densities. Depending on the form of $%
q^{\ast }(\boldsymbol{\theta })$, the expectation defined with respect to
this \textit{variational posterior }may be available in closed-form or, at
least, able to be estimated using simple Monte Carlo sampling from $q^{\ast
}(\boldsymbol{\theta }).$ INLA, on the other hand, adapts the approximation
method of Laplace to a high-dimensional setting - allied with low
dimensional deterministic integration - to produce an approximation of (\ref%
{gen_expect}). Both methods rely critically on modern techniques of
optimization, for the purpose of minimizing the `distance' between $p(%
\boldsymbol{\theta }\mathbf{|y})$ and $q^{\ast }(\boldsymbol{\theta })$ in
the case of VB, and for the purpose of producing the mode of the
high-dimensional $\mathbf{x}$, for use in the series of (nested) Laplace
approximations that underpin INLA. See \cite{ormerod2010explaining}, \cite%
{blei2017variational} and \cite{zhang2018advances}, for reviews of VB; and 
\cite{rue:martino:chopin:2009}, \cite{Rue2017}, \cite{martino2019integrated}%
, \cite{vanniekerk2019new} and \cite{wood2019simplified} for all details on
INLA.

Finally, see \cite{martin2021approximating} for a detailed outline and
comparison of \textit{all} approximate methods, including additional
discussion on `hybrid' methods that the meld features of more than one
approximate technique, with the goal of tackling multiple instances of
intractability.

\section{Postscript}

Our journey with Bayesian computation began in \textit{1763}: with a
posterior probability defined in terms of a scalar $\theta $, {whose
solution challenged} Bayes. {We now end our journey in \textit{2022}: having
referenced papers that tackle posterior distributions defined over
thousands, possibly millions of unknowns, and computational problems with a
degree of complexity \textbf{---} and scale \textbf{---} to match. Along the
way, we have seen the huge variety of imaginative computational solutions
that have been brought to bear on all such problems, over the span of 250
years. Moreover, }Bayesian computation is also beginning to confront \textbf{%
---} and adapt to \textbf{---} the reality of misspecified DGPs (%
\citealp{wangblei2019b}; \citealp{frazier2020model}; \citealp{frazier2019robust}),
and the generalizations beyond the standard
likelihood-based up-date that are evolving (\citealp{bornn2019moment}; %
\citealp{schmon2021generalized}; \citealp{frazierloss}; %
\citealp{knoblauch2022generalized}). The future of the paradigm {in the 21st
century }thus seems assured. {And with this, the 18th century Bayes (and his
loyal champion, Price) would no doubt be duly impressed!}

\bibliographystyle{imsart-nameyear}
\bibliography{Bayes_comp}

\end{document}